\documentclass[aps,superscriptaddress,showpacs,nofootinbib,eqsecnum]{revtex4}
\usepackage{amssymb}
\usepackage{amsmath}
\usepackage[dvips]{graphicx}
\usepackage{epsfig}
\usepackage{mathrsfs}

\def\d{{\rm d}}
\def\i{i}

\newcommand{\q}{\rm{q}}  
\newcommand{\uq}{{\rm u}}
\newcommand{\dq}{{\rm d}}
\newcommand{\sq}{{\rm s}}

\newcommand{\zerov}{\mathbf{0}}

\newcommand{\tauv}{\boldsymbol{\tau}}
\newcommand{\Bf}{\boldsymbol{B}}

\newcommand{\alphav}{\boldsymbol{\alpha}}

\newcommand{\p}{{\rm p}}

\newcommand{\goi}{{g_{V}^{s}}}
\newcommand{\goiv}{{g_{V}^{v}}}
\newcommand{\gi}{{\lambda_{V}^{s}}}
\newcommand{\giv}{{\lambda_{V}^{v}}}
\newcommand{\gf}{\lambda_{s}}
\newcommand{\gfv}{\lambda_{v}}
\newcommand{\I}{{\rm I}}

\begin{document}

\title{Mean field approach to flavor susceptibilities with a vector interaction}

\author{L.~Ferroni}\affiliation{Nuclear Science Division, Lawrence Berkeley National Laboratory,
1 Cyclotron Road, Berkeley, 94720}\affiliation{Institut f\"ur Theoretische Physik, Goethe-Universit\"at, 
 Max-von-Laue-Str.~1, D-60438 Frankfurt am Main, Germany}
\author{V.~Koch}\affiliation{Nuclear Science Division, Lawrence Berkeley National Laboratory,
1 Cyclotron Road, Berkeley, 94720}

\begin{abstract}
We show that flavor diagonal and off-diagonal susceptibilities of light quarks at vanishing 
chemical potential can be calculated consistently 
assuming the baryon density and isospin density dependence of QCD to be expressed by a vector-isoscalar 
and a vector-isovector coupling, respectively.
At the mean field level, their expression depends only on 
the effective medium-dependent couplings and quark thermodynamic potential. The strength of the 
couplings can be then estimated from the model using susceptibilities calculated 
in lattice QCD as an input.
\end{abstract}

\maketitle

\section{Introduction}
The recent observation of the large elliptic flow
at RHIC~\cite{Ackermann:2000tr,Voloshin:2008dg} together with the high jet energy loss, 
led to the conjecture that the matter at RHIC is strongly coupled, a nearly perfect fluid.
On the other hand, Monte Carlo simulations of QCD (Quantum Chromodynamics) on a discrete lattice
(lattice QCD)~\cite{Cheng:2008zh} suggest a quasiparticle
picture of the Quark Gluon Plasma (QGP), at least for the quarks. 
In fact, both flavor-off-diagonal susceptibilities~\cite{Koch:2005vg} and higher order
baryon number susceptibilities~\cite{Ejiri:2005wq} are consistent with vanishing correlations 
for temperatures right above the transition, $T \gtrsim 1.2T_c$~\cite{Koch:2009wk}, and appear to be 
well described by quasiparticle approaches~\cite{bluhm}. 
In addition, effective quark models with a Polyakov 
loop potential (see, for example, refs.~\cite{pnjl,crist}) 
proved to be quite successful in reproducing the thermodynamics of strongly interacting matter, 
and an even better agreement was recently found with a (2+1)-flavor Polyakov-quark-meson 
model~\cite{Schaefer:2009ui} in the mean-field approximation.
Based on these observations, it was proposed in ref.~\cite{Koch:2009wk} that the large elliptic 
flow at RHIC could be described by single-particle dynamics with a repulsive interaction. 
The same interaction would also explain the ($\sim 15\%$) deviation from the 
Stefan-Boltzmann limit of the pressure calculated in lattice QCD.

A repulsive mean-field of the vector type was also invoked
as a possible explanation for the discordant results of lattice QCD in refs.~\cite{deForcrand}, that 
appear to disfavor the existence of a critical point in the QCD phase diagram.
The authors of refs.~\cite{deForcrand} observed that the region 
of quark masses where the transition is of the first
order (for quark masses smaller than the physical ones), tends to shrink for small positive values of 
the chemical potential $\mu$. This is in contrast with model studies that support the existence of a 
critical point. In these, the first 
order region has to expand with increasing $\mu$, so that the physical quark mass point coincides 
with the critical 
line at some finite value of $T$ and $\mu$. 
As was proposed in ref.~\cite{Fukushima:2008is}, a sufficiently 
strong vector coupling may account for the shrinkage of the first order region, eventually
resulting in two critical points for a given range of (small) quark masses and in one
critical point for physical quark masses~\cite{Bowman}. 

Inspired by these ideas, we propose a scheme to estimate the strength of the vector interaction from 
lattice QCD data on flavor diagonal and off-diagonal susceptibilities. We will confine ourselves to the
light quark doublet and we will assume that, at high temperature, 
the dependence on the net-baryon density and the isospin density can be described by mean field vector-isoscalar and
vector-isovector interactions, respectively.
This is somehow in the same spirit of ref.~\cite{Kunihiro:1991qu}, where it was shown that lattice data 
on baryon number susceptibility may be described in terms
of a temperature-dependent repulsive vector coupling in the Nambu Jona-Lasinio (NJL) model. 
The vector-isovector coupling, needs also to be taken into account as its relative strength, with respect to the 
vector-isoscalar coupling, has a strong impact on the flavor off-diagonal susceptibilities~\cite{isovect}.
In our analysis, we do not make 
any specific modeling of the remnant part of the interaction (that is certainly present in QCD).
Its effect is taken into account by allowing the vector couplings, as well as   
the quark thermodynamic potential\footnote{As it will be clarified in the following, 
here we use the expression ``quark thermodynamic potential'' to indicate the thermodynamic potential of 
the quark sector at vanishing vector coupling.}, to be medium-dependent parameters.
Within these assumptions, we will consistently obtain a very simple expression 
for the flavor diagonal and off-diagonal susceptibilities at zero chemical potential.
The result can be then inverted and the difference between the vector-isoscalar and the vector-isovector
coupling can be extracted directly from lattice data using the susceptibilities evaluated in~\cite{Allton:2005gk} 
as an input.
The strength of both couplings can be then estimated considering two extreme cases that will 
allow us to obtain lower and upper bounds for these quantities as a function of the temperature.    
As we will show, the vector-isoscalar coupling is very large (repulsive interaction)
immediately below $T_{c}$ and reaches quickly much lower values above 
$T_c$, where it basically coincides with the vector-isovector coupling. 
Below $T_{c}$ also the vector-isovector coupling exhibits a similar behavior but remains always smaller 
than the vector-isoscalar coupling. 


This note is organized as follows: In Sec.~\ref{sec1} we briefly review the basic mean field results
for a massive vector field interaction and we introduce the partition function of the model. 
In Sec.~\ref{sec11} we calculate net quark number densities and susceptibilities and in Sec.~\ref{sec12} we extract 
the value of the vector couplings from lattice QCD data.

\section{The model}
\label{sec1}
In this section, we will introduce the effective partition function of our model.
To start, we will recall a few basic results of mean-field
theory in presence of a vector-isoscalar coupling $\goi$ and a vector-isovector coupling $\goiv$. 
The Lagrangian of a system of quarks with such massive vector fields is of the form
\begin{equation}
\mathcal{L}=\overline{\Psi}\left(\i \gamma^{\mu}\partial_{\mu} -m \right)\Psi-\frac{1}{4}F_{\mu \nu}F^{\mu \nu}-
\goi \overline{\Psi}\gamma^{\mu}\Psi V_{\mu}+\frac{\eta_{s}^2}{2}V^{\mu}V_{\mu} 
-\goiv \overline{\Psi}\gamma^{\mu} \tauv \Psi \Bf_{\mu}+\frac{\eta_{v}^2}{2}\Bf^{\mu}\Bf_{\mu} \;,
\label{0.0}
\end{equation} 
where $\Psi\equiv (\uq,\dq)$ is the quark field, $m$ is the quark bare mass, $\tauv\equiv (\tau^1,\tau^2,\tau^3) $ are the Pauli matrices 
and $\eta_{s}$, $\eta_{v}$ are the bare masses of the 
vector-isoscalar field $V_{\mu}$ and the vector-isovector field $\Bf_{\mu}$, respectively. 
In the Euler-Lagrange formalism we can calculate the equation of motion for the static vector fields.
By taking the average we obtain for the $0$-th component of $V_{\mu}\equiv(V_0,\boldsymbol{V})$ and 
$\Bf_{\mu}\equiv(\Bf_0,\boldsymbol{\Bf})$
\begin{eqnarray}
\langle V_0 \rangle &=&\frac{\goi}{\eta_{s}^2} \rho_{\q}\; ;\qquad  \rho_{\q} \equiv \langle \Psi^\dagger \Psi \rangle \\
\nonumber
\langle B^3_0 \rangle &=&\frac{\goiv}{\eta_{v}^2} \rho_{\I}\; ;\qquad  \rho_{\I} \equiv \langle \Psi^\dagger \tau^3 \Psi \rangle\; .
\label{0.2}
\end{eqnarray}
The average of the spatial parts $\langle \boldsymbol{V} \rangle$ and $\langle \boldsymbol{B} \rangle$ vanishes because of the 
rotational symmetry. The same is true for $\langle B^1_0 \rangle$ and $\langle B^2_0 \rangle$, because
of isospin symmetry.  
In mean field approximation, the eigenvalues of the energy of the quarks turn out to be 
(see for example~\cite{isovect,Walecka:1974qa})
\begin{eqnarray}
E_{\uq}^{\pm} (\p)&=&\frac{\goi^2}{\eta_{s}^2}  \rho_{\q} + \frac{\goiv^2}{\eta_{v}^2}  \rho_{\I} \pm \sqrt{\p^2+m^2} \\
\nonumber
E_{\dq}^{\pm} (\p)&=&\frac{\goi^2}{\eta_{s}^2}  \rho_{\q} - \frac{\goiv^2}{\eta_{v}^2}  \rho_{\I} \pm \sqrt{\p^2+m^2} \; ,
\label{0.1}
\end{eqnarray} 
with the energy for anti-$\uq$($\dq$) quarks being $-E_{\uq (\dq)}^{-} (\p)$, according 
to the ordering prescription in~\cite{Walecka:1974qa}. 
Following the standard procedure, one can  write down the partition function for this simple system.
For our discussion, we will confine ourselves to the light flavor doublet 
($\uq$ and $\dq$ quarks) and we will adopt classical (Boltzmann) statistics. 
After normal ordering one obtains
\begin{equation}
\frac{\ln Z}{{\rm V}} = \cosh \left(\frac{ \mu_{\uq}-\gi \rho_{\q}-\giv \rho_{\I}}{T}\right)\Phi_{\uq}+ 
\cosh \left(\frac{ \mu_{\dq}-\gi \rho_{\q} +\giv \rho_{\I}}{T}\right)\Phi_{\dq}
+\frac{1}{2}\frac{ \gi}{T} \rho_{\q}^2 +\frac{1}{2}\frac{ \giv}{T} \rho_{\I}^2\; ,
\label{1.1_triv}
\end{equation}
where $\mu_{\uq}$ and $\mu_{\dq}$ are the chemical potentials for the $\uq$ and $\dq$ quarks and 
$-\Phi_{\uq}T$ and $-\Phi_{\dq}T$ are the corresponding thermodynamic potentials
for $\mu_{\uq}=\mu_{\dq}=0$. In this simple case $\Phi_{\uq}$ and $\Phi_{\dq}$ are just the logarithm 
of the partition function of a gas of free quarks with mass equal to the bare mass $m$. 
At the mean field level, the couplings $\gi=\goi^2/\eta_{s}^2$ and $\giv=\goiv^2/\eta_{v}^2$ are constant.
In Eq.~(\ref{1.1_triv}), the terms $\gi \rho_{\q}^2/2T $ and $\giv \rho_{\I}^2/2T $ are sometimes referred to as  
{\em rearrangement terms}. Additional terms of this kind appear naturally to account for the energy balance 
and restore thermodynamical consistency in mean field approximation~\cite{Walecka:1975ft}.
Our analysis will be performed in the infinite volume limit. From now on, to simplify the notation, we will omit 
the system volume ${\rm V}$ in the following equations.

The Eq.~(\ref{1.1_triv}) is indeed an oversimplified partition function to describe 
a complicated theory such as QCD. Even if we want to confine ourselves to a purely phenomenological 
description through effective models, we need, at least, to include other kinds
of fields and interactions. For some purposes, it may be sufficient to add a scalar field (as in the Walecka
model~\cite{Walecka:1974qa}) or a scalar and a pseudoscalar meson 
(as in the linear sigma model with quarks~\cite{GellMann:1960np}) with their 
interaction terms. For our discussion, however, we will not need to know the exact nature of these  
interactions. We rather assume that the vector-like part of the interaction in the QCD Lagrangian 
can be isolated and treated as a mean field. In the quark sector, the effect of the rest of the interaction
is then assumed to be relegated to the vector couplings and the quark thermodynamic potential.   
Starting from Eq.~(\ref{1.1_triv}) we will consider a partition function of the form
\begin{equation}
\ln Z = \cosh \left(\frac{ \mu_{\uq}-\gi^* \rho_{\q} -\giv^* \rho_{\I}}{T}\right)\Phi^*_{\uq}+ 
\cosh \left(\frac{ \mu_{\dq}-\gi^* \rho_{\q} +\giv^* \rho_{\I}}{T}\right)\Phi^*_{\dq}
+\frac{1}{2}\frac{ \gi^*}{T} \rho_{\q}^2+\frac{1}{2}\frac{ \giv^*}{T} \rho_{\I}^2+R \; .
\label{1.1}
\end{equation} 
In Eq.~(\ref{1.1}), the couplings $\gi^*$ and $\giv^*$ are now medium-dependent quantities that  
may embody radiative corrections arising from additional pieces in the Lagrangian in Eq.~(\ref{0.0}).
Similarly, also $\Phi^*_{\uq}$ and $\Phi^*_{\dq}$ can have a non-trivial dependence on the medium.
Any kind of additional interaction, besides affecting directly the couplings and the 
quark thermodynamic potential, will in general also result in additional pieces in the partition function.
For instance, in presence of (pseudo)scalar excitations 
(besides the appearance of a thermal quark mass implicitly embodied in $\Phi^*_{\uq}$ and $\Phi^*_{\dq}$), 
the Eq.~(\ref{1.1}) should contain terms
such as the partition function of the (pseudo)scalar quasiparticles plus  
any rearrangement term needed for the thermodynamic consistency of the model\footnote{These rearrangement terms
are always present in density-dependent field theories where they  
appear in the self-energy corrections~\cite{Fuchs:1995as}. In some cases they may be interpreted as a 
consequence of the medium-dependent energy of the system in absence of quasiparticle excitations~\cite{Gorenstein:1995vm}.}.
These contributions are represented by the term $R$ in Eq.~(\ref{1.1}). As we will show below, this term does not enter in
the expression for the densities and the susceptibilities, but must be taken into account to evaluate the pressure.

Equation~(\ref{1.1}) exhibits the whole {\em explicit} dependence on $\rho_{\q}$, $\rho_{\I}$, 
$\mu_{\uq}$, $\mu_{\dq}$, $\gi^*$ and $\giv^*$ of the partition function.
In general, the quark thermodynamic potentials, the couplings $\gi^*$, $\giv^*$ and $R$ will depend on $T$ 
and on a set $\left\{\Pi_i\left(\mu_{\uq},\mu_{\dq},T \right)\right\}$ of additional 
quantities stemming from the remnant part of the interaction. 
In a thermal environment, these terms could be the scalar condensate or, as in ref.~\cite{Bowman,Mocsy:2004ab} the average fluctuations of the 
scalar mesonic fields and so on.   
Formally, our partition function can be written 
as
\begin{equation}
\ln Z\left(\mu_{\uq},\mu_{\dq},T \right) = 
\ln Z \left[\mu_{\uq},\mu_{\dq},T,\rho_{\q}\left(\mu_{\uq},\mu_{\dq},T \right),\rho_{\I}\left(\mu_{\uq},\mu_{\dq},T \right), \left\{\Pi_i
\left(\mu_{\uq},\mu_{\dq},T \right)\right\}\right] \;.
\label{stat1}
\end{equation} 
Following the general prescription in ref.~\cite{Gorenstein:1995vm} we then impose the 
thermodynamical consistency through the condition for the stationarity
\begin{equation}
\left. \frac{\partial \ln Z}{\partial \rho_{\q}}\right|_{\mu_{\uq},\mu_{\dq},T}=0\;, \qquad
 \left. \frac{\partial \ln Z}{\partial \rho_{\I}}\right|_{\mu_{\uq},\mu_{\dq},T}=0
\left.\qquad{\rm and}\qquad \frac{\partial \ln Z}{\partial \Pi_i}\right|_{\mu_{\uq},\mu_{\dq},T}=0 \;\; \forall i
\; .
\label{statcond}
\end{equation} 
Eqs.~(\ref{statcond}) need to be fulfilled by any model to 
recover the standard connection between statistical mechanics and thermodynamics\footnote{The first equation 
in Eqs.~(\ref{statcond}) should not be confused 
with the differential of the Gibbs free-energy $G$ at constant temperature and pressure: 
$\d G=\mu \d N$, (where $N$ is the number of particles and $\mu$ the chemical potential). Here 
$\rho_{\q}$ is not a {\em natural} variable of the partition function (thermodynamic potential), 
and the stationarity condition has nothing to do with a variation of the physical quantities in the system. The same
arguments hold for $\rho_{\I}.$} 
(for a detailed discussion see ref.~\cite{Gorenstein:1995vm}). In other perturbative approaches, 
such as the Optimized Perturbation Theory an analogous condition is imposed by the principle of minimal
sensitivity~\cite{pms} to determine the value of the thermal mass (see for example refs.~\cite{opt} and references therein). 
In the following we will make use of these two equations 
without making any further assumption on the functional form of $\gi^*$, $\giv^*$, $\Phi^*_{\uq}$, $\Phi^*_{\dq}$ and $R$.
To simplify the notation we recast   
the Eq.~(\ref{1.1}) in a more convenient way by using the isospin symmetry
\begin{equation}
\Phi^*_{\uq} = \Phi^*_{\dq} \equiv \Phi 
\label{1.3}
\end{equation}   
and the shorthand notations
\begin{equation}
\gf \equiv  \frac{ \gi^*}{T}\;; \qquad \gfv \equiv  \frac{ \giv^*}{T}\;;\qquad
\alpha_{\uq} \equiv \frac{ \mu_{\uq}}{T}\;; \qquad 
\alpha_{\dq} \equiv \frac{ \mu_{\dq}}{T} \; ,
\label{1.4a}
\end{equation}
leading to
\begin{equation}
\ln Z = \cosh \left(\alpha_{\uq}-\gf \rho_{\q} -\gfv \rho_{\I}\right)\Phi+ 
\cosh \left(\alpha_{\dq}-\gf \rho_{\q} +\gfv \rho_{\I}\right)\Phi+\frac{1}{2}\gf \rho_{\q}^2 +\frac{1}{2}\gfv \rho_{\I}^2+R\; .
\label{1.1b}
\end{equation} 
By further introducing the chemical potential $\mu_{\q}$ for the net quark-number 
$N_{\q} \equiv N_{\uq}+N_{\dq}$ and the chemical potential $\mu_{\I}$ for the isospin\footnote{Note that we count isospin in
units of $1$, instead of $1/2$.} 
$\I\equiv 2 \I_3=N_{\uq}-N_{\dq}$ one defines
 \begin{equation}
\alpha_{\q} \equiv \frac{ \mu_{\q}}{T}=\frac{1}{2} \left(\alpha_{\uq}+\alpha_{\dq}\right) \;; \qquad
\alpha_{\I} \equiv \frac{ \mu_{\I}}{T}=\frac{1}{2} \left(\alpha_{\uq}-\alpha_{\dq}\right) \; , 
\label{1.4}
\end{equation}
and the partition function finally reads
\begin{equation}
\ln Z = 2 \cosh \left(\alpha_{\q}-\gf \rho_{\q}\right)\cosh\left(\alpha_{\I} -\gfv \rho_{\I}\right)\Phi+
\frac{1}{2}\gf \rho_{\q}^2 +\frac{1}{2}\gfv \rho_{\I}^2+R \; .
\label{1.5}
\end{equation}

\section{Densities and susceptibilities}
\label{sec11}

The net quark-number density $\rho_{\q}$ and the isospin density $\rho_{\I}$ can be 
derived from the partition function in Eq.~(\ref{1.5}) by performing the partial derivative with respect to
$\alpha_{\q}$ and $\alpha_{\I}$, respectively. The dependence on $\alpha_{\q}$, $\alpha_{\I}$ of 
$\rho_{\q}$ and $\rho_{\I}$, and the implicit dependence on $\alpha_{\q}$, $\alpha_{\I}$ of $\gf$, $\Phi$ and $R$ can 
be ignored due to Eqs.~(\ref{statcond}). This results in the implicit expressions 
\begin{eqnarray}
\label{1.7}
\frac{\partial \ln Z}{\partial \alpha_{\q}}&=&\rho_{\q} = 
2 \sinh \left(\alpha_{\q}-\gf \rho_{\q}\right)\cosh\left(\alpha_{\I}-\gfv \rho_{\I}
\right) \Phi\; ,\\ \nonumber
\frac{\partial \ln Z}{\partial \alpha_{\I}}&=&\rho_{\I} = 
2 \cosh \left(\alpha_{\q}-\gf \rho_{\q}\right)\sinh\left(\alpha_{\I}-\gfv \rho_{\I}
\right) \Phi\; .
\end{eqnarray}
We note that, thanks to the rearrangement terms $\frac{1}{2}\gf \rho_{\q}^2$ and $\frac{1}{2}\gfv \rho_{\I}^2$, 
the first two equations in Eqs.~(\ref{statcond}) are naturally fulfilled by 
the solutions in Eq.~(\ref{1.7}) 
\begin{eqnarray}
\label{1.6do} 
\left. \frac{ \partial \ln Z}{\partial \rho_{\q}}\right|_{\mu_{\uq},\mu_{\dq},T}&=&-2 \gf \sinh \left(\alpha_{\q}-\gf \rho_{\q}\right)
\cosh\left(\alpha_{\I} -\gfv \rho_{\I} \right)\Phi+ \gf \rho_{\q}=0 \; , \\ \nonumber
\left. \frac{ \partial \ln Z}{\partial \rho_{\I}}\right|_{\mu_{\uq},\mu_{\dq},T}&=&-2 \gfv \cosh \left(\alpha_{\q}-\gf \rho_{\q}\right)
\sinh\left(\alpha_{\I} -\gfv \rho_{\I} \right)\Phi+ \gfv \rho_{\I}=0 \; .
\end{eqnarray} 
Similarly, the other rearrangement term(s) in $R$ must guarantee the validity of the third 
relation in Eq.~(\ref{statcond}). 
From Eq.~(\ref{1.1b}), one can also find the $\uq$ and $\dq$ quark net number-densities
\begin{equation}
\rho_{\uq} = \sinh \left(\alpha_{\uq}-\gf \rho_{\q}-\gfv \rho_{\I}\right)\Phi \;; \qquad
\rho_{\dq} = \sinh \left(\alpha_{\dq}-\gf \rho_{\q}+\gfv \rho_{\I}\right)\Phi \; .
\label{1.9}
\end{equation}
Because of charge conjugation- and isospin-symmetry all 
densities defined in Eqs.~(\ref{1.7}) and (\ref{1.9}) 
vanish for $\alpha_{\q}=\alpha_{\I}=0$. More generally, any odd derivative of 
the partition function vanishes.

Now that we have the densities, we can go ahead and calculate the susceptibilities.
These are obtained by further differentiating with respect to the chemical potential.
Introducing the notation $\alphav \equiv (\alpha_{\q}, \alpha_{\I})$ we define:
\begin{eqnarray}
\chi_{\q} \left(\alphav,T \right) &\equiv&  \frac{1}{T} \frac{\partial \rho_{\q}}{\partial \alpha_{\q}}= 
\frac{2}{T}  
\cosh \left(\alpha_{\q}-\gf \rho_{\q}\right)\cosh\left(\alpha_{\I} -\gfv \rho_{\I}\right)
\Phi\left(1-\gf \frac{\partial \rho_{\q}}{\partial \alpha_{\q}}-\rho_{\q}\frac{\partial \gf}{\partial \alpha_{\q}}\right)  
\nonumber \\
&-& \frac{2}{T}\sinh \left(\alpha_{\q}-\gf \rho_{\q}\right)\sinh\left(\alpha_{\I} -\gfv \rho_{\I}\right)
\Phi\left(\gfv \frac{\partial \rho_{\I}}{\partial \alpha_{\q}}+\rho_{\I}\frac{\partial \gfv}{\partial \alpha_{\q}}\right)  
\nonumber \\
&+& 2 \sinh \left(\alpha_{\q}-\gf \rho_{\q}\right)\cosh\left(\alpha_{\I}-\gfv \rho_{\I}
\right) \frac{\partial \Phi}{\partial \alpha_{\q}} \;; \nonumber \\
\chi_{\I} \left(\alphav,T \right) &\equiv& \frac{1}{T} \frac{\partial \rho_{\I}}{\partial \alpha_{\I}}=
-\frac{2}{T} \sinh \left(\alpha_{\q}-\gf \rho_{\q}\right)\sinh\left(\alpha_{\I} -\gfv \rho_{\I}\right)\Phi 
\left( \gf \frac{\partial \rho_{\q}}{\partial \alpha_{\I}} + \rho_{\q}\frac{\partial \gf}{\partial \alpha_{\I}} \right) \nonumber \\ 
&+& \frac{2}{T}\cosh \left(\alpha_{\q}-\gf \rho_{\q}\right)\cosh\left(\alpha_{\I} -\gfv \rho_{\I} \right) \Phi
\left(1-\gfv \frac{\partial \rho_{\I}}{\partial \alpha_{\I}}-\rho_{\I}\frac{\partial \gfv}{\partial \alpha_{\I}}\right)
\nonumber \\
&+& \frac{2}{T} \cosh \left(\alpha_{\q}-\gf \rho_{\q}\right)\sinh\left(\alpha_{\I}-\gfv \rho_{\I}
\right) \frac{\partial \Phi}{\partial \alpha_{\I}} \; .
\label{2.0}
\end{eqnarray}
For $\alphav=\zerov$ these two equations reduce to:
\begin{eqnarray}
\chi_{\q}  &=&  \frac{2 \Phi}{T\left(1+2\gf\Phi\right)} \; \nonumber \\
\chi_{\I}  &=& \frac{2 \Phi}{T\left(1+2\gfv\Phi\right)}\; ,
\label{2.0a}
\end{eqnarray}
where we have used the shorthand $\chi_{\q} \equiv \chi_{\q}(\zerov,T)$ and the same for $\chi_{\I}$.
The terms involving the derivatives of $\gf$, $\gfv$ and $\Phi$ with respect to 
$\alpha_{\q}$ and $\alpha_{\I}$ in Eqs.~(\ref{2.0}) do not contribute to the susceptibilities for
$\alphav=\zerov$.
Unfortunately, this is not the case for higher order susceptibilities. 
Already at the $4$-th order there are non vanishing terms involving the second derivative\footnote{For symmetry reasons, 
one expects $\Phi$, $\gf$ and $\gfv$ to depend on even powers 
of $\alpha_{\q}$ and $\alpha_I$, resulting in vanishing first 
derivatives for $\alphav=\zerov$.}
of $\Phi$, $\gf$ and $\gfv$. To evaluate these terms, one then needs to 
introduce further assumptions in the model, but this goes beyond the 
scope of our discussion. 

The diagonal and off-diagonal 
flavor susceptibilities $\chi_{\uq \uq}$ and $\chi_{\uq \dq}$ are given by 
$\chi_{\uq \uq}\left(\alphav,T \right) =  (1/T) \partial \rho_{\uq}/\partial \alpha_{\uq}$ and 
$\chi_{\uq \dq}\left(\alphav,T \right) = (1/T) \partial \rho_{\uq}/\partial \alpha_{\dq}$ which
can be conveniently obtained using (see Eqs.~(\ref{1.4}))
\begin{eqnarray}
\frac{\partial}{\partial \alpha_{\uq}}&=&\frac{1}{2}\left(\frac{\partial}{\partial \alpha_{\q}} +\frac{\partial}{\partial
\alpha_{\I}}\right) \; ; \nonumber \\
\frac{\partial}{\partial \alpha_{\dq}}&=&\frac{1}{2}\left(\frac{\partial}{\partial
\alpha_{\q}} -\frac{\partial}{\partial \alpha_{\I}}\right) \; .
\label{2.0b}
\end{eqnarray}
Therefore,
\begin{eqnarray}
\chi_{\uq \uq}\left(\alphav,T \right) &\equiv& \frac{1}{T} \frac{\partial^2 \ln Z}{\partial \alpha_{\uq}^2}=\frac{1}{4T}
\left(\frac{\partial^2}{\partial \alpha_{\q}^2} +
\frac{\partial^2}{\partial \alpha_{\I}^2}+\frac{\partial^2}{\partial \alpha_{\I}
\alpha_{\q}} \right) \ln Z  \; ;\nonumber \\
\chi_{\uq \dq}\left(\alphav,T \right) &\equiv& \frac{1}{T} \frac{\partial^2 \ln Z}{\partial \alpha_{\uq} \partial \alpha_{\dq}}=
\frac{1}{4T} \left(\frac{\partial^2}{\partial \alpha_{\q}^2} -
\frac{\partial^2}{\partial \alpha_{\I}^2} \right) \ln Z \; .
\label{2.0c}
\end{eqnarray}
For $\alphav=\zerov$ the mixed derivatives on the first equation vanishes, and
therefore:
\begin{eqnarray}
\chi_{\uq \uq} &=&  \frac{\chi_{\q}+\chi_{\I}}{4} \; ;\nonumber \\
\chi_{\uq \dq} &=&  \frac{\chi_{\q}-\chi_{\I}}{4} \; .
\label{2.0d}
\end{eqnarray}
Note that, for $\alphav=\zerov$, we have $\chi_{\uq \uq}=\chi_{\dq \dq}$  
because of isospin symmetry. From Eq.~(\ref{2.0a}) and Eq.~(\ref{2.0d}) one has: 
\begin{eqnarray}
\chi_{\uq \uq} &=&  \frac{ \Phi \left[1+\left( \gf +\gfv \right) \Phi\right]}{T\left(1+ 2 \gf \Phi\right)\left(1+ 2 \gfv \Phi\right)} \; ; \nonumber \\
\chi_{\uq \dq} &=& \frac{ \Phi^2 \left( \gfv-\gf \right)}{T\left(1+ 2 \gf \Phi\right)\left(1+ 2 \gfv \Phi\right)} \; .
\label{2.3}
\end{eqnarray}
It is worth to remark that this result, together with Eqs.~(\ref{2.0a}), does not formally depend, 
within our model assumptions, on the possible presence of other kinds of effective interaction.
Their effect is implicitly taken into account by the non-trivial dependence on the medium  
of the functions $\Phi$, $\gf$ and $\gfv$. Particularly, the presence of other (scalar) mesons 
would affect directly the value of $\Phi$ (and therefore of the susceptibilities), for instance through a 
temperature dependent effective quark mass. 
There are, however, other mesonic effects that are not included in our scheme. As shown by Eq.(\ref{2.3}), the  
off-diagonal susceptibility $\chi_{\uq \dq}$ vanishes if the isovector and
isoscalar couplings coincide (or vanish). This cannot account for the contribution of pions
described in ref.~\cite{Lombardo:2009rt}, which exist also in the absence of a vector interaction.
A more detailed discussion may be found 
e.g. in refs.~\cite{crist,Lombardo:2009rt}. 
Inverting the Eqs.~(\ref{2.3}) one finds:
\begin{equation}
\gf=\frac{1}{T \chi_{\q}}-\frac{1}{2\Phi} \;; \qquad \gfv=\frac{1}{T \chi_{\I}}-\frac{1}{2\Phi} \;. 
\label{add1}
\end{equation}
In principle, the vector couplings could be then estimated from Eqs.~(\ref{add1}) 
using lattice QCD results on flavor susceptibilities as an input.
It is, in fact, tempting to identify the pressure for $\alphav=\zerov$ with $2 \Phi T$. 
However, one should bear in mind that we are actually missing the contribution $R T$ coming 
from the unknown term $R$ in Eq.~(\ref{1.1b}). As an example, 
in the presence of additional scalar and pseudoscalar mesons as in the linear sigma 
model of ref.~\cite{Mocsy:2004ab}, the factor $R$ will account for the pressure 
generated by the mesons, plus other terms that must be introduced in order to avoid 
double counting of the fundamental degrees of freedom. 
Unlike $\chi_{\q}$ and $\chi_{\I}$, the information carried by $\Phi$ is therefore 
incomplete and cannot be immediately related to the pressure of the quarks. 
In the next section, however, we will show that is possible to
explore two limiting cases and provide upper and lower bounds for $\gf$ and $\gfv$.

\section{Vector couplings from lattice QCD}
\label{sec12}

In this section, we will provide an estimate for the temperature-dependent 
vector couplings $\gf$ and $\gfv$ starting from Eqs.~(\ref{add1}). To do that we will use, as an input, the 
susceptibilities evaluated in ref.~\cite{Allton:2005gk} in a two-flavor lattice simulation with bare quark mass 
$m/T=0.4$ at $\alphav=\zerov$. In ref.~\cite{Allton:2005gk} it was shown that $\chi_{\q}$ and $\chi_{\I}$ (and then, due to Eq.~(\ref{2.0d}),
also $\chi_{\uq \uq}$) rise steeply around $T_{c}$ and then 
reach a plateau at approximately $80\%$ of the Stefan Boltzmann (SB) limit. The off-diagonal susceptibility 
$\chi_{\uq \dq}$, instead, starts from negative values 
($\chi_{\I}$ is a bit larger than $\chi_{\q}$ at $T<T_{c}$) and then approaches zero at $T>T_{c}$.
The vanishing correlations between $\uq$ and $\dq$ quarks do not 
imply that there are no interactions. The deviation from the SB limit indicates that they are actually still 
important at $T\sim 2T_{c}$. 
From the second equation in Eqs.~(\ref{2.3}) one sees that $\chi_{\uq \dq}=0$ if $\gfv=\gf$.
The deviation from the SB limit can be then due to vector-type interactions, and/or to non-vector interactions 
reflected directly in the quark thermodynamic potential $\Phi$ (see Eqs.~(\ref{2.0a})). 

From Eqs.~(\ref{add1}), we note that the 
difference $(\gf -\gfv)$ does not depend on $\Phi$
\begin{equation}
\gf -\gfv=\frac{1}{T} \left(
\frac{1}{\chi_{\q}}-\frac{1}{\chi_{\I}}\right) \; ,
\label{add2}
\end{equation}
and can be evaluated using lattice data. 
\begin{figure}[h!]
\begin{center}
\includegraphics[width=0.8\textwidth]{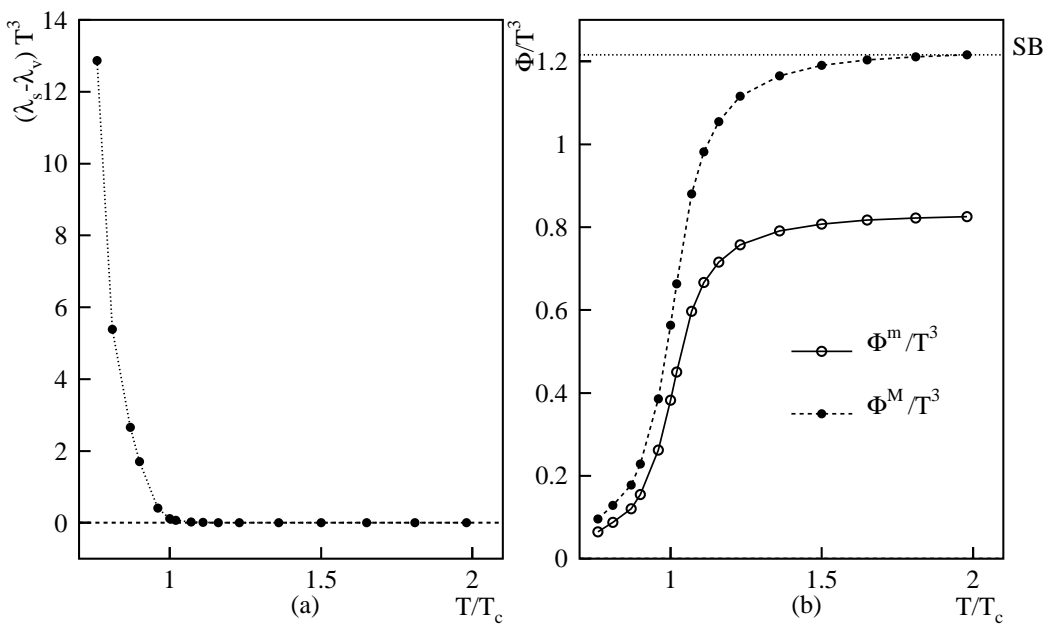}
\caption{\small{Left panel: the difference $(\gf-\gfv) T^3$ extracted from lattice 
QCD data~\cite{Allton:2005gk} on flavor diagonal and off-diagonal susceptibilities using the formula in Eq.~(\ref{add2}). 
Right panel: The assumed upper bound ($\Phi^{M}/T^3$) and lower bound ($\Phi^{m}/T^3$) for the function $\Phi/T^3$.}}
\label{f1}
\end{center}
\end{figure}
The result is shown in Fig.~\ref{f1}a. 
As one can see, the dimensionless quantity $(\gf-\gfv)T^3$ 
is large for $T/T_c<1$ and vanishes very quickly above the critical temperature. The vector-isoscalar coupling $\gf$ 
is therefore larger or equal (for $T > T_c$) to $\gfv$. This was somehow expected as in the chirally broken phase
 the coupling of the vector-isoscalar field 
(the $\omega$ meson) with the nucleons is empirically three times larger than the corresponding vector-isovector 
(the $\rho$ meson) coupling, whereas in the symmetric phase they are expected to be equal (see, for example, the discussion 
in ref.~\cite{isovect}).

As already mentioned, a unique estimate for the couplings cannot be obtained because the function $\Phi$ 
in Eqs.~(\ref{add1}) cannot be extracted from lattice data without introducing additional assumptions to quantify 
the effect of the unknown contribution to the pressure $R T$.
From Eq.~(\ref{add2}) it is clear, however, that the lower bound for $\gf$, $\gf^m$ corresponds to vanishing 
$\gfv$. 
In this case $\gf^m$ is simply given by the Eq.~(\ref{add2}) with $\gfv \equiv 0$ and the second relation in 
Eq.~(\ref{add1}) gives  
\begin{equation}
\Phi^{m} = \frac{T}{2}\chi_{\I} \; ,
\label{add3}
\end{equation}
where the superscript $m$ means that this quantity corresponds to the lower bound $\gf^m$. 
By plugging $\chi_{\I}$ from lattice QCD into the Eq.~(\ref{add3}) one can now evaluate the 
function $\Phi^m$. The result (divided by $T^3$) is the continuous curve with open circles in Fig.~\ref{f1}b. 
In this scenario the coupling $\gf$ goes to zero very rapidly above $T_{c}$ (since $\gfv=0$ the curve in Fig.~\ref{f1}a
corresponds to $\gf^mT^3$) and only interactions which are not of the vector type, such as (pseudo)scalar interactions, 
are left. The latter entirely account for the aforementioned $\sim 80\%$ deviation from the Stefan Boltzmann 
limit of the susceptibilities. As a result, the thermodynamic potential $\Phi$ is much smaller than the corresponding
SB limit for massless particles $12T^3/\pi^2$.
   
The opposite extreme case, instead, is realized by imposing that the only interaction responsible for such a deviation 
is of the vector type. As one can see from the first equation in Eqs.~(\ref{add1}), $\gf$ is maximal when $\Phi$ is maximal.
To estimate the upper bound of $\gf$ and $\gfv$ we therefore replace $\Phi\rightarrow \Phi^M$ in Eqs.~(\ref{add1}), where
$\Phi^{M} = A \Phi^{m}$ and $A$ is a constant value chosen is such a way that $\Phi^{M}(2T_{c})=$ SB limit 
(see Fig.~\ref{f1}b). This will give us the upper bounds $\gf^M$ and $\gfv^M$ corresponding to the
case where the whole deviation from the SB limit at $T>2T_{c}$ is due to the vector 
interactions\footnote{Note that if there are only interactions of the vector type the term $R$ in Eq.~(\ref{1.5}) is zero and $2 \Phi T$ is the
actual pressure of the quarks.}. This is 
quite an overestimate, as quarks are actually expected to have a thermal mass as found e.g. in the hard thermal 
loop expansion (see, for example, refs.~\cite{htl}). 
\begin{figure}[h!]
\begin{center}
\includegraphics[width=0.6\textwidth]{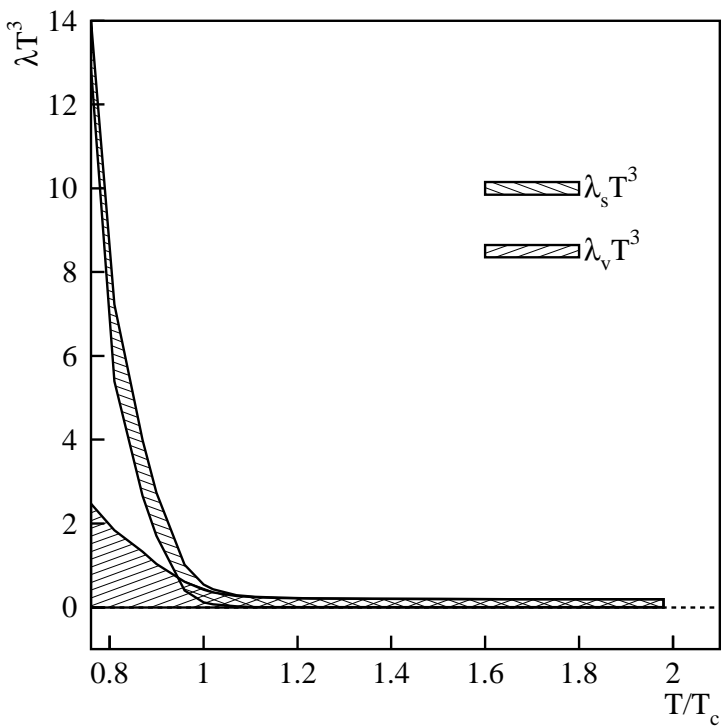}
\caption{\small{The vector-isoscalar coupling $\gf$ and the vector-isovector coupling $\gfv$ (multiplied by $T^3$) as a function of
$T/T_c$.
Shaded areas are drawn between the upper and the lower bounds representing our estimated uncertainty.}}
\label{f2}
\end{center}
\end{figure}
By construction, therefore, our estimates leave the possibility of an important vector coupling 
at high temperature (as was proposed in ref.~\cite{Koch:2009wk}) open.
To obtain tighter constraints, other (independent) flavor-related observables should 
be considered. One possibility, could be higher order susceptibilities. 
This, however, requires the introduction of additional assumptions as mentioned in Sec.~\ref{sec11}. 
The admissible values for $\gf$ and $\gfv$ are represented by the shaded areas in Fig.~\ref{f2}.  
As one can see, the dimensionless quantity $\gf T^3$ 
is large (repulsive interaction) for $T/T_c<1$ and reaches quickly much lower 
values above the critical temperature. A similar behavior is shown also by $\gfv$ that, however, is smaller 
than $\gf$ at low $T$. 

As mentioned in the introduction, the effect of a vector-isoscalar coupling was also accounted for in a recent 
work~\cite{Fukushima:2008is} to describe
the shrinkage of the first order region in the $m_{\uq\dq}, m_{\sq}$ plane (where $m_{\uq\dq}$ is the light quark mass 
and $m_{\sq}$ the strange quark mass) at small chemical potential $\mu$. 
This effect was observed in lattice QCD~\cite{deForcrand} and led to the conclusion that the existence 
of a critical point (at least at small $\mu/T$) is unlikely.
The author of ref.~\cite{Fukushima:2008is} was able to reproduce this  behavior in the Polyakov loop extended 
NJL model (PNJL) by introducing a repulsive vector-vector interaction of the isoscalar type. 
For a sufficiently strong  vector coupling\footnote{Notice that the author of ref.~\cite{Fukushima:2008is} uses a different 
definition for the vector coupling $G_V\equiv \gi/2$.} 
($\gi\sim 0.8 \lambda$, where $\lambda$ is the scalar coupling) the first order region 
initially tends to shrink as $\mu$ is increased to small positive values, and, after reaching its minimum size, 
it tends to expand as $\mu$ is further increased. 

The method proposed in this note could be, in principle, used to extract the value of the vector-isoscalar 
coupling in coincidence with the critical line $T=T_{c}$ and 
$m_{\uq\dq}^c=m_{\sq}^c$, where $m_{\uq\dq}^c$ and $m_{\sq}^c$ are the chiral critical masses.
To do that, however, further ingredients are needed. First, our treatment should be extended to three flavors. 
Second, also the scalar coupling should be extracted from lattice data in a consistent fashion.
In effective quarks models such as the NJL model, in fact, the scalar coupling $\lambda$ depends on the cutoff 
and the value commonly used in literature~\cite{Hatsuda:1994pi,buballa}  
ranges from $\lambda \sim 5$~GeV$^{-2}$ to $\lambda \sim 10$~GeV$^{-2}$. To provide a rough comparison 
(with all the aforementioned caveats in mind) from our analysis we obtain,
assuming $T_c=0.17$~GeV, $4$~GeV$^{-2}\lesssim \gi \lesssim 19$~GeV$^{-2}$. At this level, therefore, no conclusions
could be drawn in this respect.

\section{Summary}
\label{sec13}

We proposed a method to estimate the strength of the vector interaction from 
lattice QCD data on flavor diagonal and off-diagonal susceptibilities.
Our fundamental assumption is that the baryon density and isospin density dependence 
of QCD can be described as a mean field by a vector-isoscalar and a vector-isovector coupling, respectively.
Imposing thermodynamical consistency conditions, we showed that in our framework flavor susceptibilities
can be parametrized by the medium-dependent effective couplings $\gf$ (vector-isoscalar), $\gfv$ (vector-isovector)
and by the quark thermodynamic potential $\Phi$. 
The effect of other kinds of interaction, like e.g. scalar or pseudoscalar,
affects the value of these quantities, but not the form of the result.  
Inverting our relations, we have shown that the difference $(\gf-\gfv)$ depends only on $\chi_{\q}$ and $\chi_{\I}$ (or,
equivalently, $\chi_{\uq \uq}$ and $\chi_{\uq \dq}$) and can be then extracted 
from the model using the susceptibilities from lattice QCD~\cite{Allton:2005gk} as an input.  
We found that $(\gf-\gfv)$ is quite large below $T_{c}$ and approaches zero very rapidly above $T_{c}$. 

Unlike their difference, the value of the vector couplings alone depends on the details
of the remnant part of the interaction. Therefore, they cannot be determined directly from lattice QCD without 
introducing further assumptions.
We have then estimated upper and lower bounds for $\gf$ and $\gfv$ (using again lattice data as an input) 
considering two quite extreme scenarios. The first (yielding the lower bounds), where the effect of the non-vector part of
the interaction (such as scalar or pseudoscalar) is maximum and the second (yielding the upper bounds) 
assuming that the only interaction left at $T>2T_{c}$ is of the vector type. 
Immediately below $T_{c}$, our results suggest 
a strong $\gf$ that rapidly approaches lower values at $T>T_{c}$.
A similar behavior is observed for $\gfv$ that, however, remains always smaller than $\gf$ at low temperature.

In future works this method should be refined in order to provide tighter bounds for the 
possible values of the couplings. This could be done considering other
flavor-dependent observables from QCD (such as higher order susceptibilities) and/or 
motivating realistic assumptions to estimate the bounds. Our analysis should be also extended to three flavors. 
This could be helpful to understand if the strength of the vector coupling $\gf$ is large enough to produce 
a change of sign of the curvature of the critical surface in the $m_{\uq\dq}, m_{\sq},\mu$ space as 
proposed in ref.~\cite{Fukushima:2008is}.

\section{Acknowledgments:}
This work was supported  by the Director, Office of Energy
Research, Office of High Energy and Nuclear Physics, Divisions of
Nuclear Physics, of the U.S. Department of Energy under Contract No.
DE-AC02-05CH11231 and by the Helmholtz International Center for FAIR within the
framework of the LOEWE program (Landesoffensive zur Entwicklung Wissenschaftlich-\"Okonomischer Exzellenz) launched by the
State of Hesse.


\end{document}